\documentclass[12pt]{iopart}
\usepackage{graphicx}
\usepackage{lineno}

\begin{document}

\title[]{Hybrid Kinetic–MHD Simulations of Drift-Orbit Effects onto the Stability and Non-Linear Dynamics of Runaway Electron Beams}

\author{Shi-Jie Liu$^1$\footnote{corresponding author},
  Hao-Wei Zhang$^1$,
  Hannes Bergstroem$^1$,
  Matthias Hoelzl$^{1,2}$,
  the JOREK team\footnote{see the author list of Ref.~\cite{Hoelzl_2024} for the JOREK Team.}
  }

\address{$^1$Max Planck Institute for Plasma Physics, 85748 Garching b. M., Germany\\
$^2$Department of Physics and Astronomy, Chalmers University of Technology, Göteborg, SE-41296, Sweden}

\ead{\mailto{shi-jie.liu@ipp.mpg.de}}
\vspace{10pt}
\begin{indented}
\item[]
\end{indented}

\begin{abstract}
During tokamak disruptions, the Ohmic current may be replaced by a non-inductive runaway electron (RE) current, affecting resistive stability. Previous analytical and numerical studies suggest that, in the linear phase, the presence of REs acts destabilizing for tearing modes (TM) compared to a scenario with Ohmic current. In the non-linear regime, this translates to larger saturation amplitudes; in particular, for small $\Delta^\prime$, the saturated TM island width is approximately 1.5 times larger than that in the corresponding Ohmic scenario. These results, however, are based on the assumption of zero drift orbit deviation from the magnetic flux surfaces, which corresponds to the low RE energy limit. This work investigates the importance of this kinetic effect, i.e., linear and non-linear TM dynamics are studied in RE beams with different RE energies, providing a clear picture of the FOW effects. We use a hybrid fluid-kinetic model in the 3D non-linear magnetohydrodynamic (MHD) code JOREK, treating REs kinetically with a full-f Monte-Carlo approach in self-consistent interaction with the MHD mode dynamics. The study shows that the presence of REs modifies the characteristics of the instability in several ways. Firstly, we find that the major-radial displacement of drift orbits from flux surfaces induces an m=1 perturbation to the equilibrium current introducing additional mode coupling of m,n instabilities with $m\pm1,n$ modes, where m and n refer to the poloidal and toroidal mode numbers, respectively. Secondly, we find that increasing RE energy has a stabilizing effect on the MHD modes since REs cannot support narrow current sheets on rational flux surfaces due to the drift orbit displacements, reducing both the linear growth rate and the non-linear saturation level. This counteracts the destabilizing effect that REs have on TMs in the low energy limit. For the scenario investigated, the stabilizing effects dominate over the additional mode coupling such that the development of stochastic magnetic regions is reduced with increasing RE energies, leading to lower radial particle transport. Overall, our findings indicate that FOW effects can substantially alter MHD stability and non-linear dynamics of RE beams.

\end{abstract}
\vspace{2pc}
\noindent{\it Keywords}: Magnetohydrodynamics, runaway electron, disruption, mode coupling, transport
%
%
%
%
%

\section{Introduction}

Tokamak disruptions pose a critical challenge for the safe operation of future fusion devices, such as ITER, primarily due to the generation of runaway electron (RE) beams\cite{Salewski_2025,Breizman_2019}. During the disruption process, the pre-disruption Ohmic current can be rapidly converted to RE current. The presence of such a non-thermal current\cite{Lvovskiy_2020} fundamentally alters the stability properties of the post-disruption plasma.

Typically, tokamak disruptions\cite{R.D.Gill_1993,Bandyopadhyay_2025,10.1063/1.3703327} consist of two phases: the thermal quench (TQ) and the current quench (CQ). During the TQ phase, there is a rapid and near complete loss of plasma thermal energy, significantly increasing the resistivity of the plasma. Following this, the Ohmic current, carried by the thermal electrons, dissipates during the CQ phase. As the resistivity of the plasma increases, a toroidal electric field is induced, continuously accelerating electrons. Due to this acceleration, a RE plateau can form during the CQ\cite{M.N.Rosenbluth_1997}. In this stage, the plasma is typically at low temperature and density, while a significant fraction of the current is carried by highly energetic REs. The low collisionality of the REs stops the current decay in this phase. Such RE beams pose a serious threat to plasma-facing components (PFC), as the relativistic particles may be lost in a highly localized manner, leading to severe damage. For this reason, a strategy for safe termination has been developed to prevent intolerable heat fluxes to the wall, which is called RE beam benign termination\cite{Reux_2021,Paz-Soldan_2021}. It refers to a scenario where the RE population is flushed out of the plasma in a very short timescale, following the induction of a global MHD event that stochastizes the magnetic field and distributes the particle flux over a wide surface area on the PFCs\cite{Bandaru_2021, Zimmermann_2026}. Understanding the interaction between RE and MHD instabilities in the post-disruption phase is therefore of critical importance. 

Previous studies\cite{10.1063/1.2817016,10.1063/5.0018559} have investigated both the linear and non-linear regimes of resistive MHD modes in the presence of REs. In the linear phase, REs enhance the growth rate of TMs at high plasma resistivities compared to the corresponding Ohmic case. Furthermore, in the non-linear regime, RE-driven currents can lead to larger saturation widths of the magnetic islands. However, these analyses are based on simplified assumptions, such as treating REs as moving at the speed of light and neglecting FOW effects. In realistic disruption scenarios, REs exhibit a broad energy distribution\cite{Lvovskiy_2020}, and FOW effects can play a significant role in determining both equilibrium and stability properties. In particular\cite{10.1063/5.0165240}, REs exhibit notable deviations between particle orbits and flux surfaces, and also induce a Grad-Shafranov like shift of the flux surfaces, both primarily contributed by RE curvature drift.

JOREK is a massively parallel, fully implicit, non-linear extended MHD simulation code developed to study large-scale plasma instabilities in realistic tokamak geometries~\cite{Hoelzl_2021,Hoelzl_2024}. In simulations of REs, a fluid description is often employed, where REs are modeled as a distinct fluid population\cite{Bandaru_2021,10.1063/5.0213962}. However, such an approach does not resolve the energy of the particles, needed to accurately model their drift motion perpendicular to the magnetic field lines. In recent years, the development and application of hybrid kinetic–fluid models within JOREK have been presented in \cite{Bergström_2025,Liu_2026}, using full orbit and guiding-center (GC) orbit treatments, respectively. This approach advances RE particle motion by time-dependent orbital equations describing the RE population by a full-f particle-in-cell approach. To achieve a self-consistent treatment of the RE mutual interaction with the background plasma, the kinetic REs are coupled to the fluid via a pressure coupling scheme. Using this framework, we investigate the impact of FOW effects, primarily arising from the curvature drift of REs, on resistive TM by varying the RE energy.

We show that the presence of REs modifies the equilibrium, leading to a Grad–Shafranov shift and induces a (1,0) perturbation of the equilibrium current profile, thereby enhancing poloidal mode coupling and more complex non-linear dynamics \cite{Yu_2019}. For example, this coupling could lead to the generation of a (3,1) sideband from the dominant (2,1) mode. Furthermore, increasing RE energy results in a stabilizing effect on TMs, reducing the linear growth rate and altering the non-linear dynamics. As a consequence, the radial transport of REs is found to be reduced in the scenario considered here, i.e., the stabilizing effects dominate over the enhanced mode coupling in the non-linear dynamics. The particle transport evaluation in different RE beam scenarios is quantified through several indicators, including the Chirikov parameter, the radial diffusion coefficients and the connection length. Addressing these questions is essential for assessing the evolution and termination of RE beams. 

The remainder of this paper is organized as follows. In section~\ref{sec2}, we introduce the hybrid model used in this work. In section~\ref{sec3}, we analyze the impact of REs on plasma equilibrium, including FOW effect. In section~\ref{sec4}, we investigate the FOW effects of REs on resistive TM and the resulting RE transport. Finally, in section~\ref{sec5}, we summarize key findings and discuss directions for future studies.

\section{Hybrid Kinetic-MHD Model}\label{sec2}
The hybrid kinetic–MHD model used in this work has been developed and implemented in JOREK in\cite{Bergström_2025,Liu_2026} where the detailed formulation and numerical implementation are described. Here, we briefly summarize the aspects relevant to the present study. 

The relativistic GC model follows the formulation described in~\cite{10.1063/1.2773702}:
\begin{eqnarray}
\dot{\mathbf{X}}=&\frac{p_\parallel^\ast}{\gamma_r m_0}\frac{\mathbf{B}^\ast}{B_\parallel^\ast}+\frac{\mathbf{b}}{ B_\parallel^\ast}\ \times\left(p_\parallel\frac{\partial{\mathbf{b}}}{\partial t}-q\mathbf{E}+\frac{\mu}{\gamma_r}\nabla B\right)\nonumber,\nonumber\\
{\dot{p}}_\parallel=&\frac{\mathbf{B}^\ast}{B_\parallel^\ast}\cdot\left(q\mathbf{E}-p_\parallel\frac{\partial{\mathbf{b}}}{\partial t}-\frac{\mu}{\gamma_r}\nabla B\right)\nonumber,
\label{eq17}
\end{eqnarray}
where
\begin{eqnarray}
B_\parallel^\ast=&\mathbf{B}^\ast\cdot\mathbf{b}\nonumber,\\
\gamma_r=&\sqrt{1+\frac{p_\parallel^2}{(m_0 c)^2}+\frac{2\mu  B}{m_0 c^2}}\label{eq18},
\end{eqnarray}
and $\mathbf{X},\ p_\parallel,\ \mu=\frac{p_\bot^2}{2 m_0 B}$ represent the GC location, the momentum parallel to the magnetic field, and the magnetic moment of the particle, respectively. $\gamma_r$ is the RE relativistic factor. $\mathbf{E}$ and $\mathbf{B}$ denote the electric and magnetic fields, and the effective magnetic field is given as $\mathbf{B}^\ast=p_\parallel \nabla \times \mathbf{b} +q \mathbf{B}$, q is the charge, $m_0$ is the rest mass of the particle, $\mathbf{b}$ is the unit vector of magnetic field. The speed of light in vacuum is given by $c$. The GC equations include several drift terms, such as the E×B, $\nabla B$, and curvature drifts. In this work, the pitch parameter is fixed at pitch=0.999, corresponding to $p_\parallel/p=0.999$, while the RE energy is varied to investigate the FOW effect. Both curvature and $\nabla B$ drifts increase with energy. In the evaluation of the RE current, the contribution associated with the $\nabla B$ drift is canceled by the corresponding RE magnetization current\cite{10.1063/5.0165240}. So that the FOW effects from REs on the fluid is mainly determined by the curvature drift.

With thermal electrons and ions as well as REs, extended MHD equations can be derived. The modified MHD equations are given below, with only the relevant equations shown here.

The momentum equation is 

\begin{eqnarray}
    \label{momentum_pressure2}
\rho \left(\frac{\partial\mathbf{u}}{\partial t} + \mathbf{u} \cdot \nabla \mathbf{u} \right)
& = & - \sigma_r \mathbf{E}_\parallel + \mathbf{J} \times \mathbf{B} - \nabla p - \nabla \cdot \mathbf{\Pi} \nonumber \\
&   & - (\mathcal{P}_{r,\parallel}-\mathcal{P}_{r,\perp})\mathbf{\kappa}
      - \nabla\mathcal{P}_{r,\perp}.
\end{eqnarray}

The pressure is denoted by $p$, $ \mathbf{J}$ is the current density. The $\sigma_r=-e n_r$ is RE charge density. The term $\mathbf{\Pi}$ represents the viscous stress tensor. The perpendicular and parallel pressure of REs are defined as:

\begin{eqnarray}
    \mathcal{P}_{r,\perp} =\frac{1}{2} \int d^3v \gamma_r m_0 v^2_\perp f, \nonumber\\
    \mathcal{P}_{r,\parallel}= \int d^3v \gamma_r m_0 v^2_{\parallel} f.
\end{eqnarray}

RE distribution is described by the function $f(\mathbf{X}, \mathbf{v}, t)=\sum_{j=1}^{N} w_j \delta(\mathbf{X}-\mathbf{X}_j(t))\delta(\mathbf{v}-\mathbf{v}_j(t))$, where $w_j, \mathbf{v}_j$ are the weight, and velocity for the j-th particle, and N is the total number of markers. $v_\parallel$ and $v_\perp$ represent the components of the RE velocity parallel and perpendicular to the magnetic field, respectively. Since the RE current is evolved kinetically, it is excluded from the Ohmic response. Consequently, the Ohm's law is modified to
\begin{eqnarray}
    \mathbf{E}=-\mathbf{u}\times\mathbf{B}+\eta(\mathbf{J}-\mathbf{J}_r),
\end{eqnarray}
where $\mathbf{J}_r$ represents the RE current density.

\section{Finite-Orbit-Width Effects on the Plasma Equilibrium}\label{sec3}
\begin{figure}
        \centering
        \includegraphics[width=12cm]{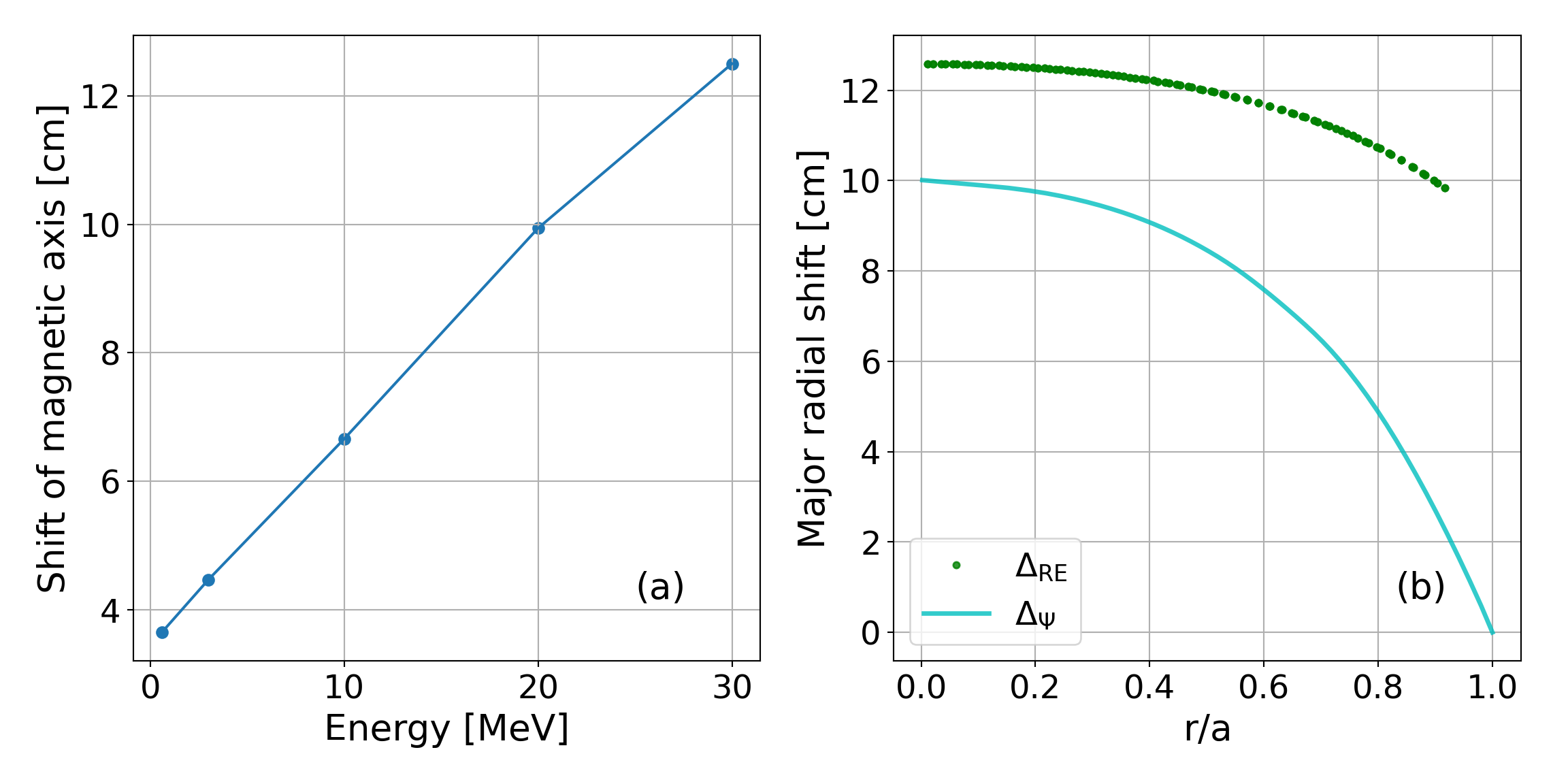}
        \caption{Equilibrium: (a) the magnetic axis shift as a function of RE energy; (b) radial shift profiles of 20 MeV REs, with the blue line indicating the flux-surface shift and the green dots representing the RE orbit shift.}
        \label{figure1}
\end{figure} 
\begin{figure}
        \centering
        \includegraphics[width=12cm]{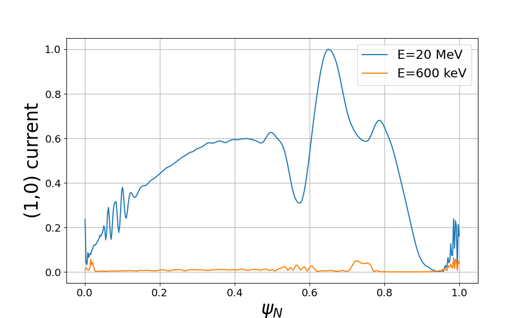}
        \caption{$(1,0)$ component of RE current in different RE energy beams, with blue line representing 20 MeV and orange line representing 600 keV.}
        \label{figure2}
\end{figure} 

In the post-disruption phase, the plasma is typically characterized by low temperature and low pressure, and tends toward a force-free equilibrium $(\mathbf{J} \times \mathbf{B} \approx 0)$. However, when a significant population of high-energy REs is present, they can provide an effective pressure term, breaking the force-free assumption $(\mathbf{J} \times \mathbf{B} = \mathcal{P}_{r,\parallel} \mathbf{\kappa})$, in which $\mathbf{\kappa}$ is the curvature vector. This leads to a G-S radial shift of the magnetic flux surfaces\cite{10.1063/5.0165240}. This effect has been analyzed via our hybrid model, and has been benchmarked against analytical results.

In this section, a cold plasma with $I_p=I_{RE}=0.6\ \mathrm{MA}$, and a $R_0=10.0\ \mathrm{m},\ a=1.3\ \mathrm{m}$ circular geometry is assumed. The safety factor profile is monotonic, with the on-axis value $q_0=1.2$ and an edge value $q_a=4.0$. The equilibrium is primarily unstable to the (m,n)=(2,1) tearing mode. The mono-energetic runaway electrons are traced through relativistic GC equations. After introducing the runaway current and turning off the current source term, the background current will naturally dissipate due to the high resistivity, and a RE beam is formed. To obtain plasmas with different RE energies, the desired RE energy is prescribed during the initialization of the electron distribution, after which the plasma is allowed to evolve self-consistently. Here, the 2D equilibrium calculations shown in this section are based on fixed-boundary equilibria. The free-boundary formulation is employed for the subsequent 3D simulations.

Figure~\ref{figure1} (a) shows the magnetic axis shift as a function of RE energy, owing to the change in force balance, and (b) radial shift profiles of 20 MeV REs, with the blue line indicating the flux-surface shift and the green dots representing the RE orbit shift. A clear deviation of several centimeters between the RE orbits and the magnetic flux surfaces is observed. And the deviation of RE orbits from flux surfaces breaks the symmetry of the equilibrium on the magnetic-flux-surface coordinates, generating a (1,0) component that leads to m=1 sidebands, which could be observed in Figure~\ref{figure2}. Figure~\ref{figure2} compares the (1,0) component of the RE current for an RE energy of 600 keV and 20 MeV as a function of the normalized poloidal flux, $\psi_N=({\psi-\psi_{\mathrm{axis}})}/({\psi_{\mathrm{bnd}}-\psi_{\mathrm{axis}}})$, where $\psi_{\mathrm{axis}}$ and $\psi_{\mathrm{bnd}}$ denote the poloidal flux at the magnetic axis and boundary, respectively. Here, the RE energy in this work refers to the total particle energy, including the electron rest energy. Clearly, the drift orbit deviations from flux surfaces, which increase with the RE energy, cause a substantial (1,0) perturbation to the MHD equilibrium that facilitates enhanced mode coupling.

\section{Finite-Orbit-Width Effects on MHD Instability}\label{sec4}

\subsection{Linear Properties}
\begin{figure}
        \centering
        \includegraphics[width=12cm]{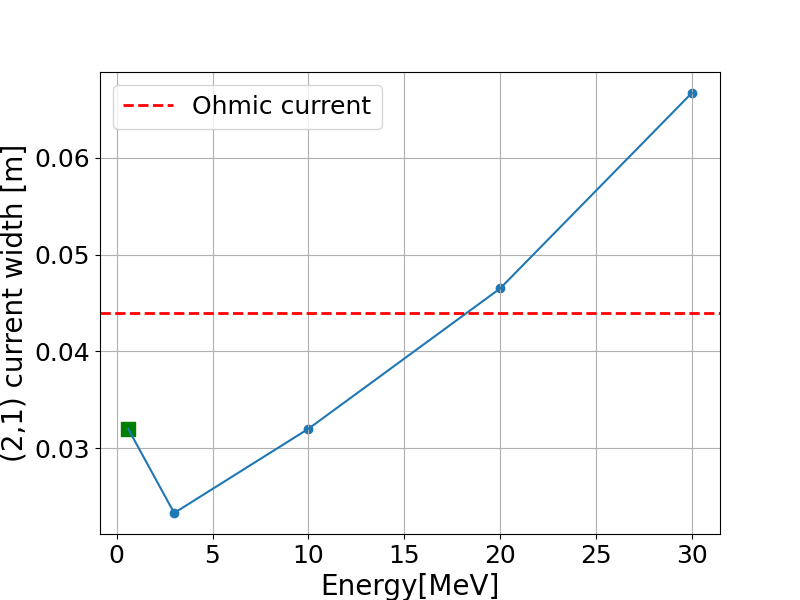}
        \caption{Width of (2,1) current layer as a function of RE energy.}
        \label{fig3}
\end{figure} 
\begin{figure}
        \centering
        \includegraphics[width=12cm]{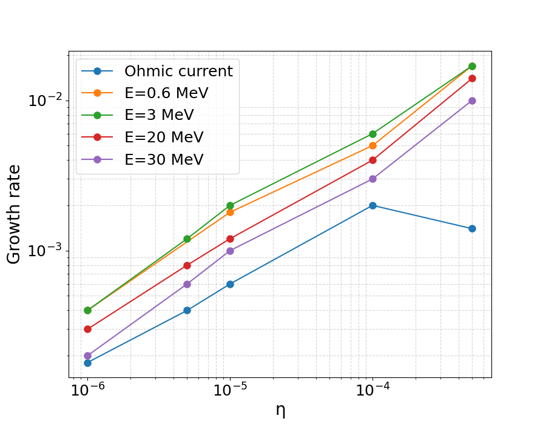}
        \caption{Growth rates of (2,1) mode, with different color corresponding to different energies, while the blue line represents the Ohmic current plasma.}
        \label{fig4}
\end{figure} 
\begin{figure}
        \centering
        \includegraphics[width=12cm]{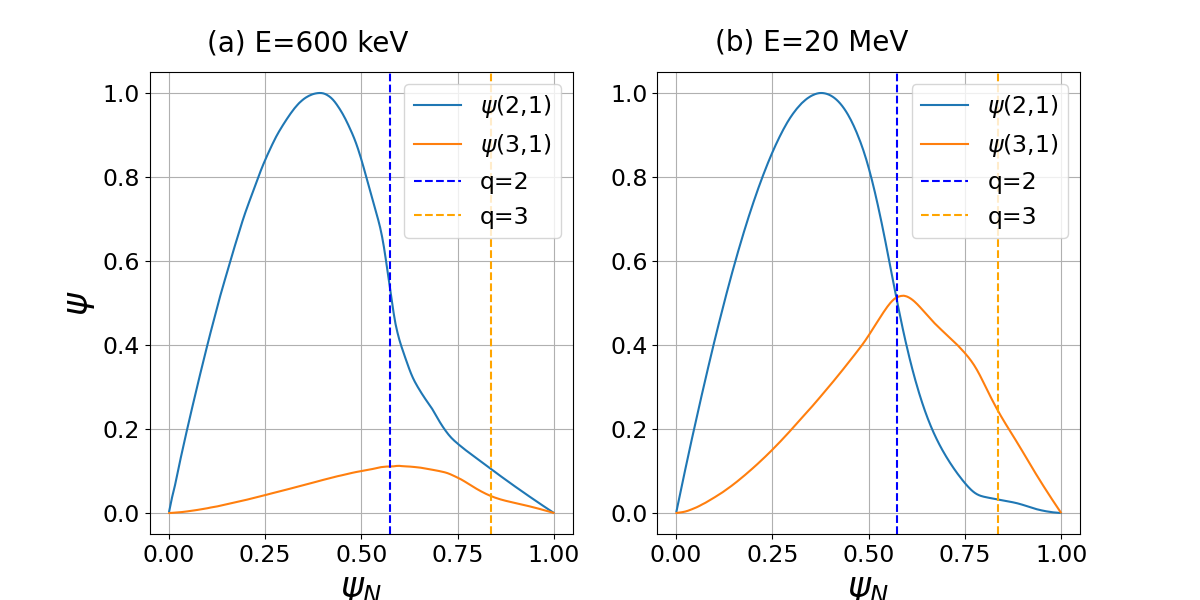}
        \caption{(2,1) (3,1) $\psi$ structures in (a) 600 keV and (b) 20 MeV RE beams. }
        \label{fig5}
\end{figure} 
In section~\ref{sec3}, we notice that the FOW effects becomes non-negligible when RE energy reaches several MeV. In\cite{10.1063/1.2817016,10.1063/5.0018559}, the differences between RE-driven MHD instability and the Ohmic current case are discussed, but neither of them includes the effect of the FOW effects of REs. In\cite{10.1063/5.0018559}, an expression for the sublayer within the resistive layer in RE-driven MHD is given,
\begin{eqnarray}
    \delta_{RE}=\gamma v_A/(k_c v_{RE})   
\end{eqnarray}
which is inversely proportional to the RE velocity, and much narrower than in the Ohmic case: $\delta_{RE}\ll\delta_{ohmic}$ with $\delta_{ohmic}=\gamma^{1/4} \eta^{1/4}k_c^{-1/2}$. Here, $\gamma$ is the growth rate, $k_c=nq^\prime/(Rq)$, $v_A$ is Alfvén speed, and $\eta$ is the resistivity.

The resistive layer width sets the characteristic radial scale of non-ideal MHD effects near the rational surface. In this framework, the current sheet associated with the (2,1) mode can be regarded as a physical manifestation of the resistive layer. In Figure~\ref{fig3}, the width of the (2,1) current density structure is shown as a function of RE energy. For comparison, a subluminal RE-beam case is also included (green square), corresponding to a total energy of 600 keV and a velocity of approximately half the speed of light $c$. In this scenario, the resistive-layer width is larger than that in the near-relativistic case (3 MeV, corresponding to approximately 0.99$c$). When RE velocity approaches light speed, the layer width reaches its minimum. This behavior is consistent with the prediction in $\delta_{RE}$. However, as the RE energy increases further, the resistive-layer width broadens again, which is a direct result of the FOW effects. The current carrying layer width is increased at high RE energies again by the deviation of RE orbits from the flux surfaces.

Consequently, for ultra-relativistic particles, the linear growth rate of the (2,1) mode is investigated. As shown in Figure \ref{fig4}, the linear growth rate is shown, normalized to the Alfvén time.
The dependence of the growth rate on the RE energy is examined for different values of resistivity. At a given resistivity of the background companion plasma, the growth rate of the TM decreases monotonically with increasing RE energy, getting close to the corresponding growth rate of a purely Ohmic plasma at high RE energies. This behavior is consistent with the corresponding increase in resistive-layer width.

Another FOW effects discussed in Sec. \ref{sec3} is the poloidal mode coupling, which may modify the overall mode structure and growth rate. To further examine the possible influence of FOW-induced coupling, the structures of the (2,1) and (3,1) modes are compared in Fig. \ref{fig5} for 600 keV and 20 MeV RE beams. It is observed that in the high-energy case, the fundamental (2,1) TM considered here develops a pronounced (3,1) sideband structure, indicating enhanced toroidal mode coupling. This additional coupling is a direct consequence of the (1,0) component of the RE current density at high RE energies.

\subsection{Non-Linear Dynamics}
\begin{figure}
        \centering
        \includegraphics[width=12cm]{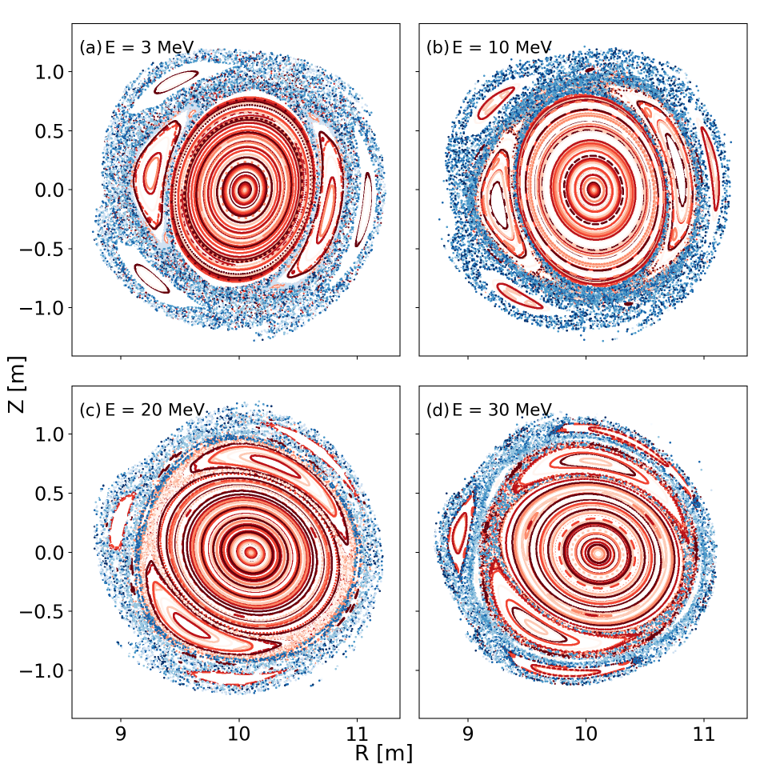}
        \caption{Magnetic field Poincaré-plot at $\phi=0$ RZ-plane for different RE beams.(a-d): 3 MeV, 10 MeV, 20 MeV, 30 MeV. The plots are obtained by tracing field lines. Red points denote confined lines, while blue points represent escaping lines.}
        \label{fig6}
\end{figure} 
\begin{figure}
        \centering
        \includegraphics[width=12cm]{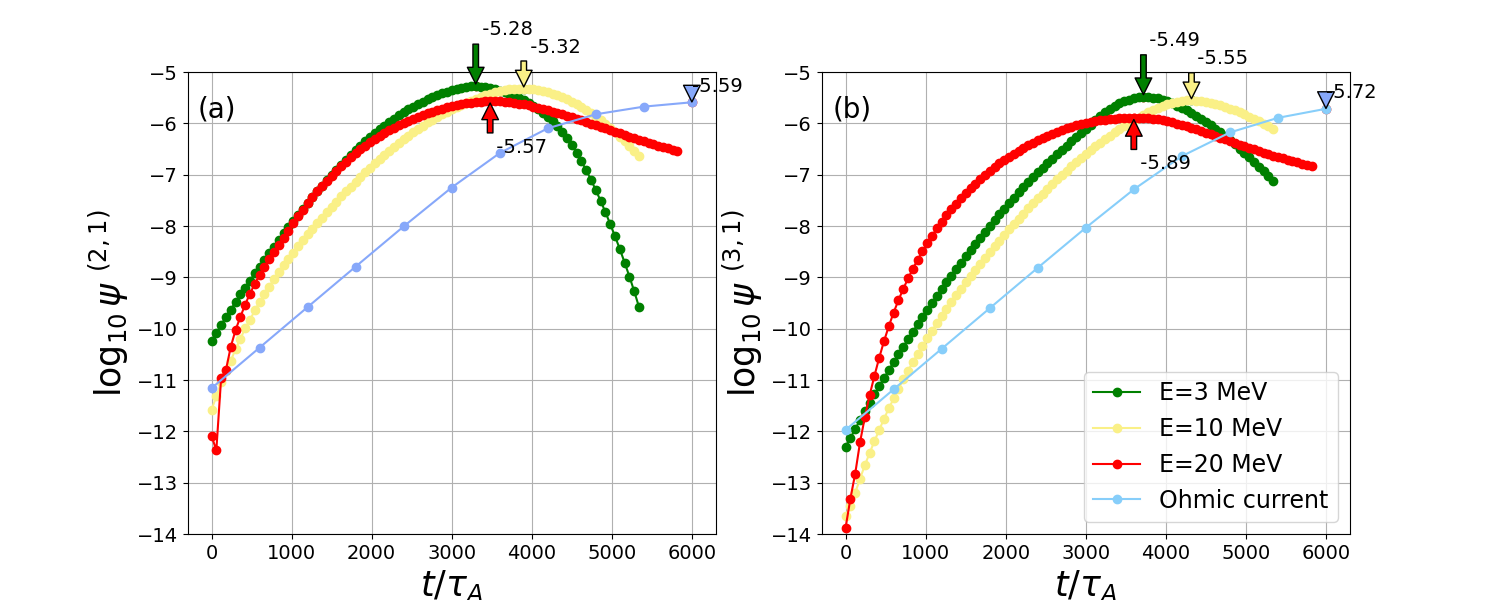}
        \caption{$\psi$ evolution of (a) (2,1) mode and (b) (3,1) mode, with different color corresponding to different energies, while the blue line represents the Ohmic current plasma. The saturation amplitudes are indicated by text labels.}
        \label{fig7}
\end{figure} 
As the system evolves from the linear to the non-linear regime, different modes continue to grow and their mutual interactions may lead to island overlap, thereby increasing the complexity of the magnetic topology. In particular, the onset of stochastic magnetic fields can enhance particle transport. For REs, this transport is of particular relevance, since RE losses during the CQ may counteract avalanche multiplication, on one hand. On the other hand, stochastic losses are critical for the termination, since the magnetic topology and the time evolution of it can strongly affect the RE wetted area and thus influence the peak heat loads and risk of damage.

Here, the non-linear evolution of the TM in RE plasma is studied for the scenarios already introduced earlier in this article, which were already assessed from a linear stability point of view. Figure \ref{fig6} shows four Poincaré plots corresponding to four RE energies of (a) 3 MeV, (b) 10 MeV, (c) 20 MeV, and (d) 30 MeV, illustrating the resulting MHD instabilities. Each plot is taken at the time when the instability reaches its maximum amplitude. In these plots, blue lines denote escaping field lines, while red lines represent confined field lines. The Poincaré sections show a dominant (2,1) island together with a pronounced (3,1) island. A significant spatial overlap between these modes leads to the formation of a stochastic magnetic region.

To better understand the origin of differences, it is useful to analyze the dynamics of the underlying individual modes.
In Figure \ref{fig7}, we compare the time evolution of the $(2,1)$ and $(3,1)$ modes for different RE beam energies, with the corresponding saturation amplitudes labeled beside the curves. We evaluate the $\psi_{m,n}$ components at their corresponding rational surfaces, such that the amplitudes provide a proxy for the (m,n) island size, which is proportional to the square-root of this $\psi_{m,n}$ component.

The time is normalized to the Alfvén time $\tau_A\approx 0.2\ \mu s$. The Ohmic case is also included for comparison. For the $(2,1)$ mode, a slight reduction in the saturation amplitude can be observed with increasing RE energy. The behavior of the $(3,1)$ mode is more subtle. In the 20 MeV case, its initial growth is enhanced substantially compared to the lower energy cases due to mode coupling. However, owing to the enhanced drift orbit stabilization effect at higher RE energies, the saturated $(3,1)$ mode energy still exhibits a decreasing trend as the RE energy increases. Overall, for the scenario considered here, the non-linear MHD activity is suppressed due to enhanced FOW effects of REs. It is worth noting again that the kinetic RE effects have several different influences on linear and non-linear mode activity, which may be counter-acting. This makes it hard to draw general conclusions for the non-linear MHD dynamics in RE beams with different RE energies thus emphasizing how important self-consistent high-fidelity modeling capabilities are. An overall stabilizing effect like we see here may be responsible for the observation of quasi-stationary MHD activity in RE beams\cite{Sommariva_2024}. It could also play an important role for the development of the explosive onset of RE termination by initially delaying dynamics compared to an Ohmic plasma.

Let's take a look at the further evolution from the peak amplitude of the perturbations. The stochastic fields lead to a partial loss of REs from the beam. The corresponding RE current is back-converted into thermal current in this process. Also for the thermal current, there is some redistribution in the stochastic field (acting effectively like a hyper-resistivity), that can change the current profile and tends to reduce current gradients at the rational surfaces. Furthermore, the thermal currents are subject to resistive decay, which happens on relatively short time scales given the cold temperatures of the RE companion plasma. Furthermore, a slight reheating of the beam is taking place when the current has become thermal again due to Ohmic heating, thus reducing the resistivity and TM growth rates. Finally, the linear growth rate of TMs is lower in an Ohmic plasma compared to the same plasma with REs carrying the current (as discussed earlier in this article). All these processes lead to a reduction of the TM linear growth rates and thus a drop of the perturbation amplitudes after the initial peaking.

\subsection{Transport Evaluation}
\begin{figure}
        \centering
        \includegraphics[width=12cm]{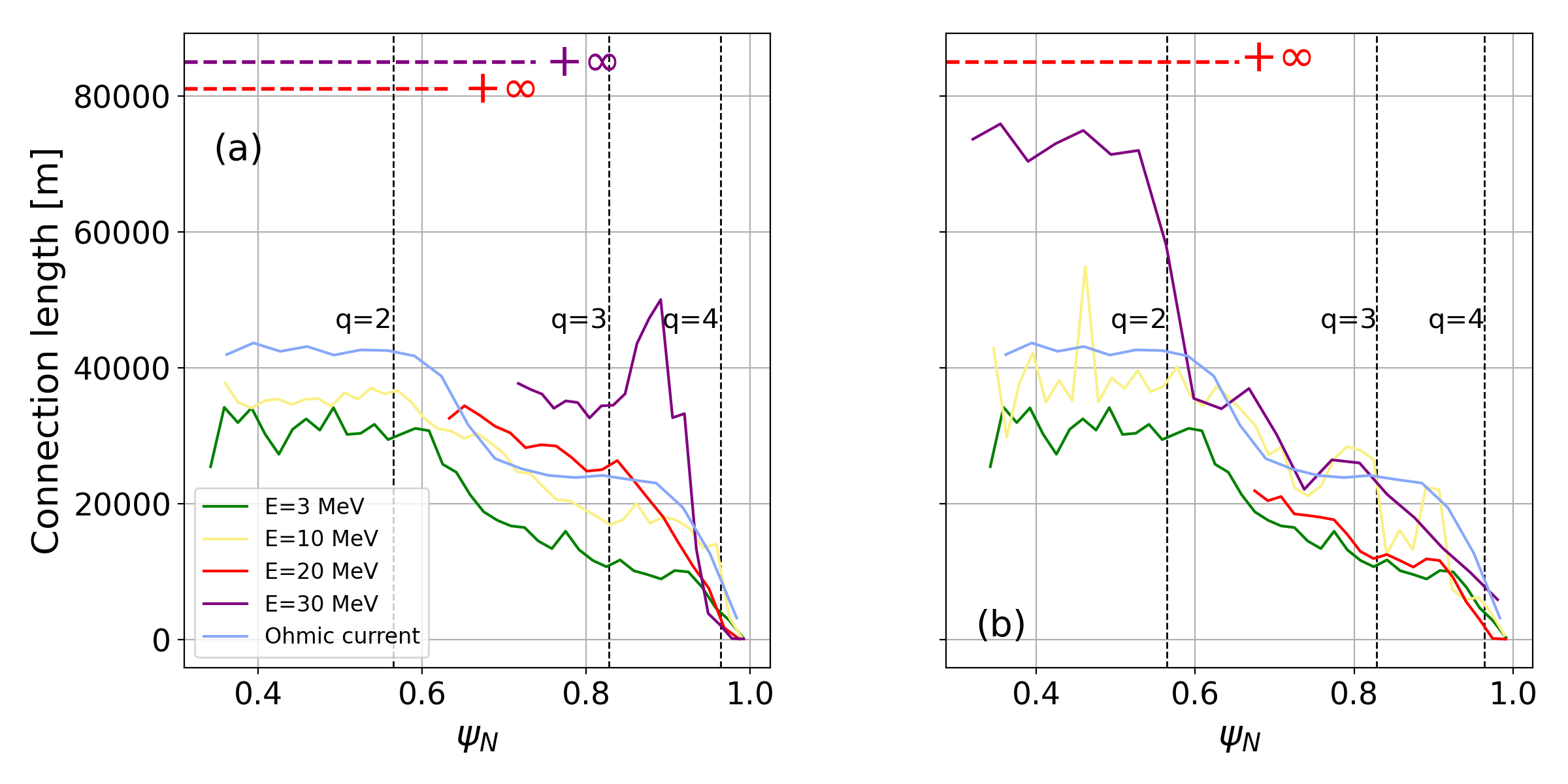}
        \caption{Connection length profiles for different RE beams, (a) field lines; (b) markers corresponding to RE beam energies.}
        \label{fig8}
\end{figure} 
\begin{figure}
        \centering
        \includegraphics[width=12cm]{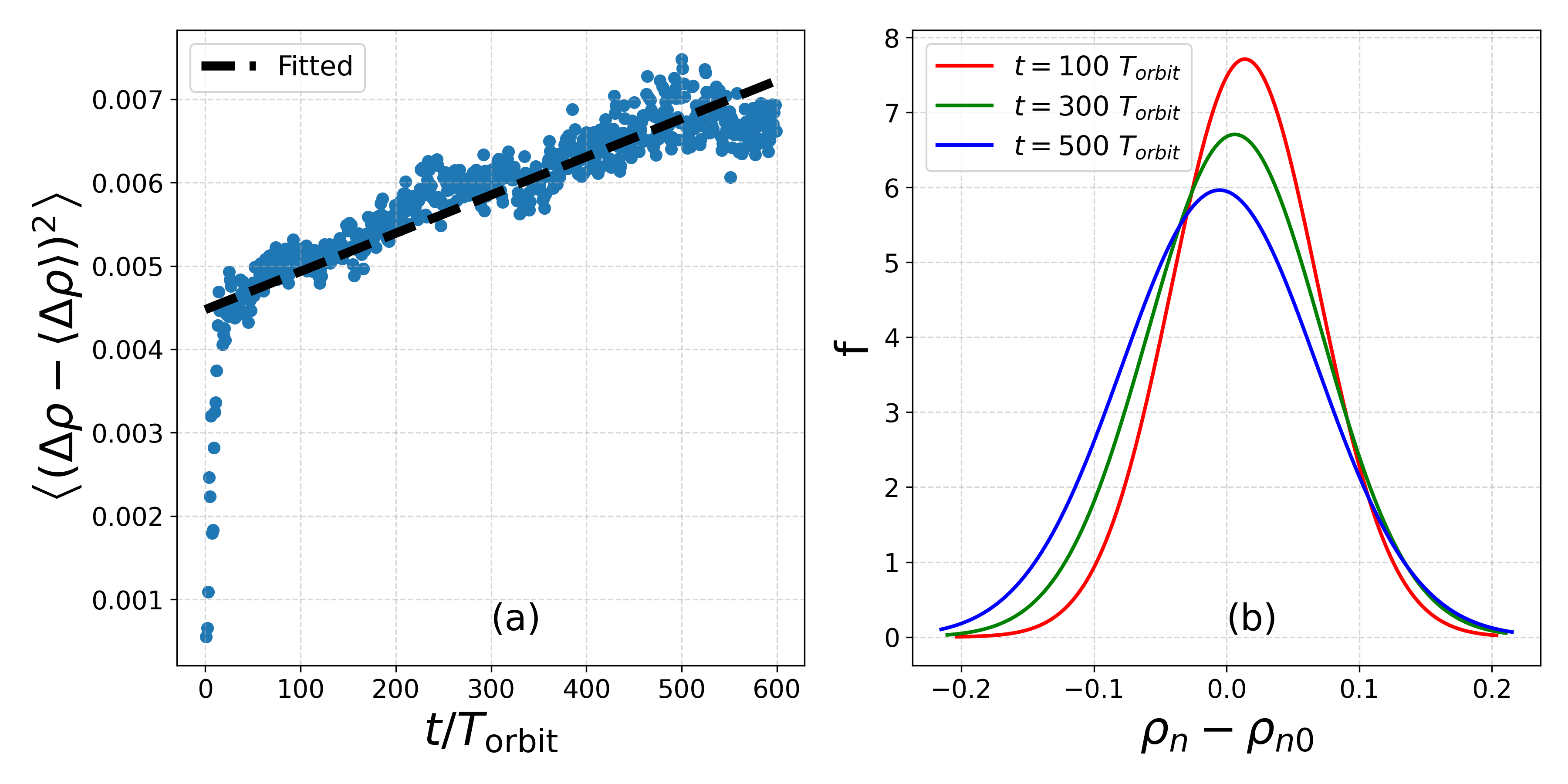}
        \caption{(a) The mean square displacement evolution and (b) the distribution of the radial displacement at several times in the Gaussian fits.}
        \label{fig9}
\end{figure} 
\begin{figure}
        \centering
        \includegraphics[width=12cm]{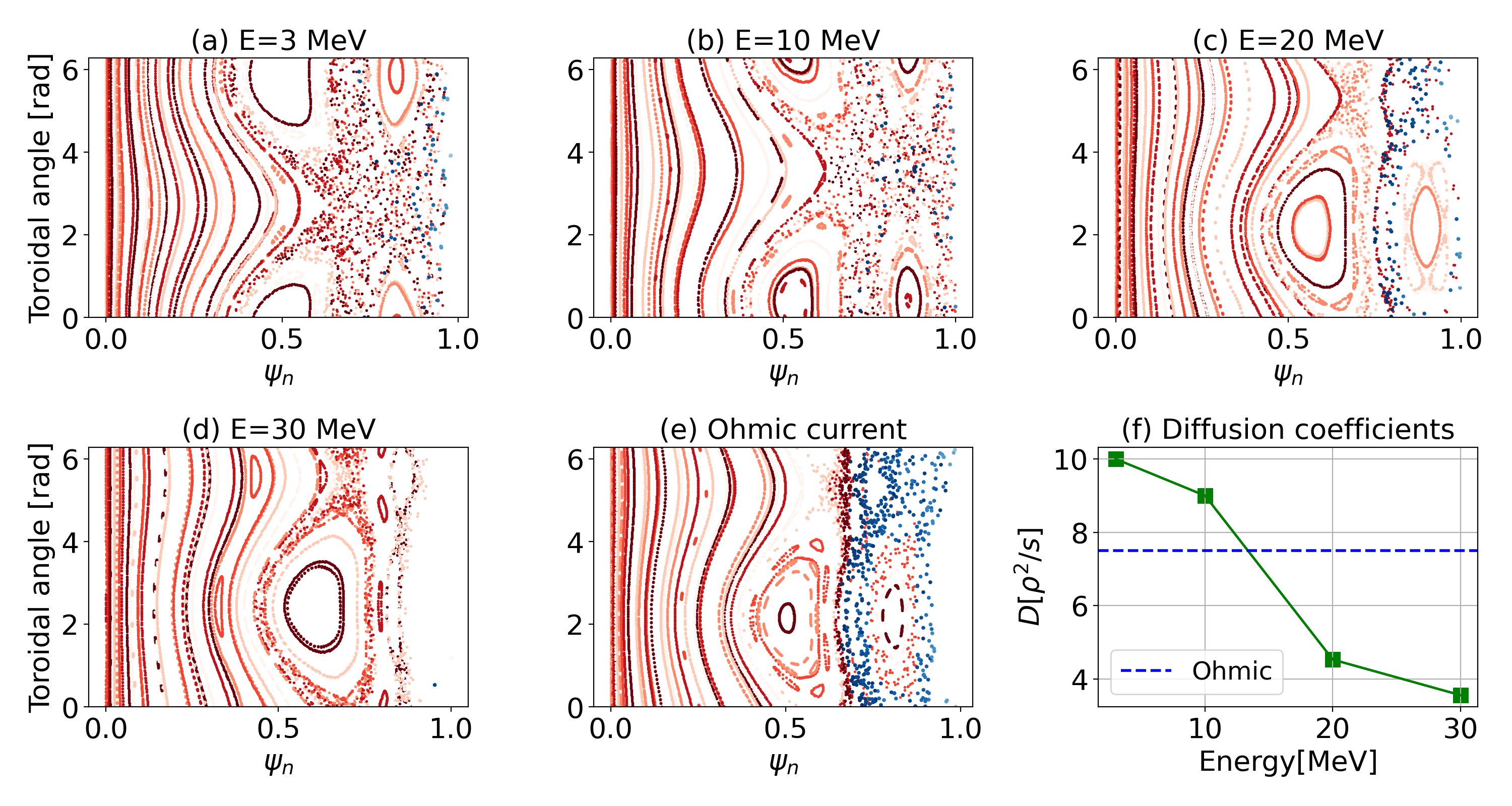}
        \caption{Field-line trajectories in $(\psi_n,\phi)$ space for different RE beam energies, (a-d) correspond to 3 MeV, 10 MeV, 20 MeV and 30 MeV, respectively, (e) represents the Ohmic current plasma case; (f) the diffusion coefficients D as a function of RE energy.}
        \label{fig10}
\end{figure} 
\begin{figure}
        \centering
        \includegraphics[width=12cm]{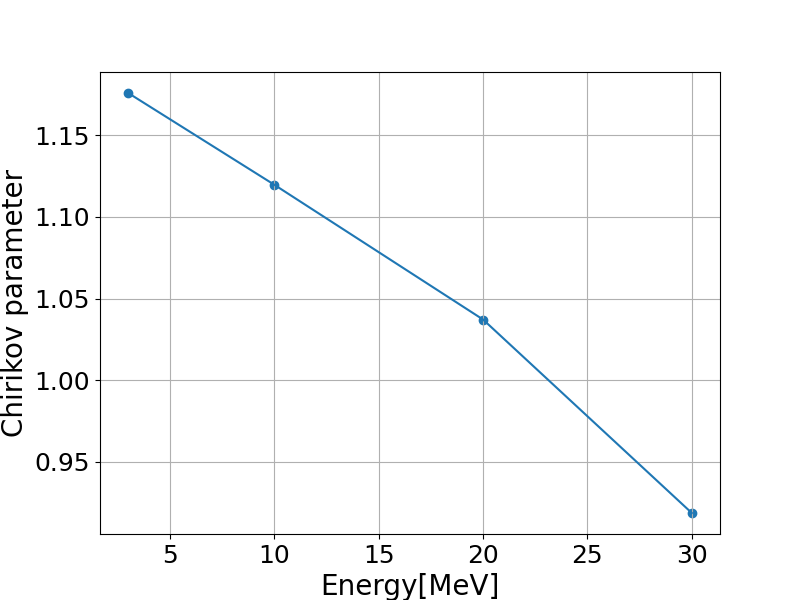}
        \caption{Chirikov parameter as a function of RE energy.}
        \label{fig11}
\end{figure} 

Poincare plots provide a good indication for the magnetic topology, but they are usually not easy to interpret in a quantitative manner. The degree of stochastization can vary greatly and thus also the influence on RE transport. The point density in regions where stochastic field lines eventually reach the computational boundary can provide a first indication regarding the degree of stochastization. However, analysis of the connection length and field line diffusivity can provide a much cleaner picture. Consequently, in this subsection, we employ several diagnostics, including connection length analysis, transport coefficient calculation, and an assessment of the Chirikov parameter, to get more precise insights into the magnetic topology for the non-linear scenarios discussed already in the previous subsection. These quantities allow us to systematically evaluate the extent of stochastic regions and their dependence on RE energy. Higher-energy electrons generally possess larger orbit widths and different resonance conditions with magnetic fluctuations. Another aspect is the degree of stochastization ``seen'' by the REs. The gyration and FOW excursions cause REs to sample different regions of the magnetic perturbation spectrum. The resulting averaging effect reduces the coherence between particle trajectories and individual magnetic field lines, thereby modifying stochastic transport\cite{Särkimäki_2020,10.1063/1.860431}, usually leading to reduced stochastic transport for highly energetic REs when considering the same magnetic topology (since the particle velocity becomes nearly independent of the energy in the relativistic regime). Therefore, when characterizing transport in stochastic magnetic fields, FOW effects must be taken into account in addition to field-line-based measures.

Let's first take a quantitative look by considering the connection length. The connection length is defined as the distance traveled by a magnetic field line or a particle before intersecting the computational boundary (considered to be the location of material structures). It provides an estimate of particle transport out of the domain. Two distinct ways can be used to quantify the connection length: one based on magnetic field-line tracing and the other based on particle-orbit tracing. The particle-based connection lengths additionally incorporate FOW effects. 

Figure \ref{fig8}(a) shows the connection length evaluated from field-line tracing. Here, the dashed horizontal lines denote regions where field lines remain confined throughout the maximum field-line tracing length considered. These regions correspond to completely intact magnetic surfaces and are therefore assigned an effectively infinite connection length. A clear increase in connection length is observed with increasing RE energy, suggesting that RE loss rates would be decreasing.

Also particle connection length analysis taking into account the different RE beam energies is shown in Figure \ref{fig8}(b). In this case, although the dependence on energy is not monotonic, the overall trend remains a reduction of transport with increasing RE energy.

An additional way of looking at the influence of magnetic topology on RE transport is via effective radial transport coefficients. This sheds more light on the local transport properties. In particular, magnetic perturbations and stochastic field structures can significantly increase cross-field RE transport. The efficiency of such transport processes is strongly dependent on particle energy \cite{Särkimäki_2020,10.1063/1.5135588}. Various transport models have been developed to describe these effects, ranging from quasilinear approaches \cite{PhysRevLett.40.38} to test particle simulations \cite{Särkimäki_2016,Papp_Drevlak_Pokol_Fülöp_2015}. Following the approach of Ref.~\cite{Särkimäki_2016}, we estimate the radial diffusion coefficient corresponding to the radial particle transport in the presence of stochastic magnetic fields. It is derived under the idealized assumption of homogeneous transport and unbounded spatial domains. Here, the transport coefficient is calculated using the normalized radial coordinate $\rho = \sqrt{({\psi-\psi_{\mathrm{axis}})}/({\psi_{\mathrm{bnd}}-\psi_{\mathrm{axis}}})}$. In this framework, if the initial particle distribution is localized and can be represented by a Dirac delta function, i.e., $f(\rho,t=0)=\delta(\rho_0-\rho)$, the subsequent time evolution of the distribution follows a Gaussian profile due to the combined effects of transport. The analytical solution can be written as
\begin{eqnarray}
f(\rho,\tau)=\frac{1}{\sqrt{4\pi D\tau}}\exp{[-\frac{(\rho-\rho_0-K\tau)^2}{4D\tau}]},
\end{eqnarray}
where K denotes the convection coefficient and D is the diffusion coefficient. Based on this solution, the function fitting method is used, in which the distribution function is directly fitted to the Gaussian form above to determine K and D. In the presence of magnetic islands and edge effects, the particle displacement statistics deviate from this idealized Gaussian spreading. We exclude these particles which belong to a non-Gaussian transport regime, whereas our analysis assumes Gaussian diffusive statistics of stochastic background. Therefore, in our case, particles which are strongly affected by the magnetic island region or boundary transport are removed from the analysis such that the transport coefficients leave out the effect of the local population in island remnants. In the analysis, 1000 markers are initialized at each radial location but are uniformly distributed in the toroidal direction, and tracked for 0.2 ms, during which they complete approximately 600 poloidal transits if not lost earlier. The final state of the particle distribution is then used to compute the transport properties, and a spatial average is performed to obtain a global estimate of the radial stochastic transport for each scenario. For consistency, the marker energy is taken to be the same as the corresponding RE energy in the hybrid fluid-kinetic simulation. To verify whether the particle transport exhibits diffusive behavior, the mean square displacement (MSD) is analyzed,
\begin{eqnarray}
\mathrm{MSD}
=
\left\langle
\left(
\Delta \rho - \left\langle \Delta \rho \right\rangle
\right)^2
\right\rangle .
\end{eqnarray}
For a diffusive process, the MSD is expected to increase linearly with time, whereas deviations from linear scaling would indicate anomalous transport.
As shown in Figure \ref{fig9}(a), the MSD increases approximately linearly with time, consistent with diffusive transport, and the initial deviation from the linear scaling arises from finite-time trajectory correlations. The correlation time $( \tau_c )$ characterizes the time scale over which particle trajectories remain correlated before entering the diffusive regime. The diffusion coefficients are therefore evaluated in the regime $( t \gg \tau_c )$, where the MSD exhibits linear scaling. Figure \ref{fig9}(b) shows the distribution of the radial displacement at several times in the Gaussian fits. Therefore, the transport coefficients obtained from Gaussian fitting can be regarded as meaningful effective transport coefficients, and are shown in Figure \ref{fig10}(f). Figure \ref{fig10} shows field-line trajectories in $(\psi_n,\phi)$ space for different RE beam energies. Panels (a-d) correspond to 3 MeV, 10 MeV, 20 MeV and 30 MeV, respectively, whereas panel (e) represents the Ohmic current plasma case. The diffusion coefficients D, calculated from particle tracing at the corresponding RE beam energies, are annotated in the panel (f). Consistently with earlier observations regarding the connection length, the radial diffusion coefficients are decreasing with increasing RE energy.

Another way of getting insights into the magnetic field topology for the different simulations is by analyzing the size of the main islands. Stochastization naturally involves also high order resonances, but the main contributors in our scenario are the (2,1) and (3,1) magnetic islands. The primary cause of magnetic field stochasticity in a tokamak is the overlap of such magnetic islands. These islands are typically generated by instabilities such as tearing modes at different rational surfaces. When the islands grow large enough, they begin to overlap. According to the Chirikov criterion\cite{10.1063/5.0173642,10.1063/1.2907163}, this overlap leads to chaotic magnetic field lines, breaking the closed magnetic surfaces and resulting in a stochastic structure. If the Chirikov value is larger than 1, it indicates significant overlap and the onset of stochastic magnetic field lines. It can be expressed as \cite{Liu_2016}:
\begin{eqnarray}
S=\frac{(w_1+w_2)/2}{|r_2-r_1|},
\end{eqnarray}
where $w_1,\ w_2$ are the widths of two adjacent magnetic islands, and $|r_2-r_1|$ is the distance between the two islands. Figure \ref{fig11} shows the Chirikov parameter calculated for the (2,1) and (3,1) islands as a function of RE energy. The monotonic decrease of the Chirikov parameter with energy shows a reduction in the degree of magnetic stochasticity, again in line with earlier observations of connection length and transport coefficients. Both the Chirikov parameter and particle transport suggest that higher RE energies exhibit a stabilizing effect on the MHD instability. 

\section{Conclusion}\label{sec5}
In this work, we investigated the FOW effects of REs on resistive TM dynamics using the hybrid kinetic–MHD model implemented in JOREK. By treating runaway electrons kinetically through relativistic guiding-center equations while evolving the background plasma with reduced MHD equations, we analyzed both equilibrium modifications and instability dynamics over a wide range of RE energies. 

The simulations demonstrate that FOW effects of RE beams play an important role in the linear and non-linear dynamics of MHD activity that are not captured by RE fluid models. The drift orbit deviation of the RE orbits from flux surfaces introduces a large (1,0) perturbation to the current that strongly enhances the coupling of (m,n) modes to (m$\pm1$,n) modes at high RE energies. On the other hand, these drift orbit deviations also prevent the formation of very localized current structures on the rational surfaces thus acting stabilizing on the MHD modes. So, while the presence of REs acts very destabilizing for TMs in the limit of particles following the flux surfaces (no drift-orbit deviations; RE fluid limit), kinetic corrections act stabilizing in contrast, thus driving the TM growth rates in an RE beam with high RE energies closer to the ones in an Ohmic plasma again.

In the non-linear regime, higher RE energies lead to reduced magnetic-island saturation amplitudes and weaker stochastic structures. So, for the considered scenario, the stabilizing effects dominate over the influence of the enhanced mode coupling. The resulting particle transport analysis and the Chirikov parameter both show a decreasing trend with RE energy increasing. These results indicate that energetic RE beams can stabilize resistive MHD activity and in turn reduce RE stochastic transport.

Overall, this study highlights that FOW effects of REs cannot be neglected in realistic disruption scenarios. Furthermore, the different effects on growth rates and mode coupling lead to a non-trivial overall effect onto the dynamics of different scenarios, thus emphasizing the need for self-consistent high-fidelity modeling. Understanding the interaction between MHD instabilities and runaway electrons, including these effects, is important for predicting the CQ evolution and (benign) termination of RE beams in future fusion devices such as ITER. Future work will extend the present analysis toward more realistic RE beam termination conditions. Self-consistent MHD–RE interaction and FOW effects will be investigated under these conditions.

\section*{Acknowledgments\label{Acknowledgements}}
Part of this work has been carried out within the framework of the EUROfusion Consortium, 
funded by the European Union via the Euratom Research and Training Programme 
(Grant Agreement No 101052200 — EUROfusion). Views and opinions expressed 
are however those of the author(s) only and do not necessarily reflect those of the 
European Union or the European Commission. Neither the European Union nor the 
European Commission can be held responsible for them. The authors gratefully acknowledge computing time on the Viper system operated by MPCDF.
\section*{Declaration of interests\label{Declaration of interests}}
The authors report no conflict of interest.

\clearpage

\bibliographystyle{iopart-num}

\bibliography{instructions}

\end{document}